\documentclass{desyproc}
\usepackage{graphicx,amsmath,amssymb}

\usepackage{mciteplus}

\begin{document}
\title{Saturation in nuclei}

\author{{\slshape T. Lappi}\\[1ex]
Department of Physics, 
 P.O. Box 35, 
40014 University of Jyv\"askyl\"a, 
Finland  and \\
Helsinki Institute of Physics, 
 P.O. Box 64, 
00014 University of Helsinki, 
Finland \\
}

\contribID{lappi\_tuomas}

\desyproc{DESY-PROC-2009-xx}
\acronym{EDS'09} 
\doi  


\newcommand{\ud}{\, \mathrm{d}}
\newcommand{\uc}{{\mathrm{c}}}
\newcommand{\ul}{{\mathrm{L}}}
\newcommand{\intd}{\int \!}
\newcommand{\nc}{{N_\mathrm{c}}}
\newcommand{\nf}{{N_\mathrm{F}}}
\newcommand{\nosum}[1]{\textrm{ (no sum over } #1 )}
\newcommand{\na}{\, :\!}
\newcommand{\nb}{\!: \,}
\newcommand{\cf}{C_\mathrm{F}}
\newcommand{\ca}{C_\mathrm{A}}
\newcommand{\df}{d_\mathrm{F}}
\newcommand{\da}{d_\mathrm{A}}
\newcommand{\nr}[1]{(\ref{#1})}
\newcommand{\dadj}{D_{\mathrm{adj}}}
\newcommand{\ra}{R_A}

\newcommand{\gev}{\ \textrm{GeV}}
\newcommand{\fm}{\ \textrm{fm}}
\newcommand{\ls}{\Lambda_\mathrm{s}}
\newcommand{\qs}{Q_\mathrm{s}}
\newcommand{\qsa}{Q_\mathrm{sA}}
\newcommand{\qsp}{Q_{\mathrm{s}p}}
\newcommand{\lqcd}{\Lambda_{\mathrm{QCD}}}
\newcommand{\as}{\alpha_{\mathrm{s}}}

\newcommand{\cN}{\mathcal{N}}
\newcommand{\cNt}{\widetilde{\mathcal{N}}}

\newcommand{\raa}{R_\mathrm{AA}}
\newcommand{\rpa}{R_\mathrm{pA}}

\newcommand{\fig}{Fig.~}
\newcommand{\figs}{Figs.~}
\newcommand{\eq}{Eq.~}
\newcommand{\se}{Sec.~}
\newcommand{\eqs}{Eqs.~}

\newcommand{\rt}{\mathbf{r}_T}
\newcommand{\bt}{\mathbf{b}_T}
\newcommand{\xt}{\mathbf{x}_T}
\newcommand{\yt}{\mathbf{y}_T}
\newcommand{\itt}{\mathbf{i}_T}
\newcommand{\jt}{\mathbf{j}_T}
\newcommand{\pt}{{\mathbf{p}_T}}
\newcommand{\ptt}{p_T} 
\newcommand{\qt}{\mathbf{q}_T}
\newcommand{\kt}{\mathbf{k}_T}
\newcommand{\lt}{\mathbf{l}_T}
\newcommand{\ut}{\mathbf{u}_T}
\newcommand{\vt}{\mathbf{v}_T}
\newcommand{\nabt}{\boldsymbol{\nabla}_T}

\maketitle

\begin{abstract}
This talk discusses some recent studies of gluon saturation in nuclei. 
We stress the connection between the initial condition in heavy ion 
collisions and observables in deep inelastic scattering (DIS).
 The dominant degree of freedom in the small 
$x$ nuclear wavefunction is a nonperturbatively strong classical gluon field,
which determines the initial condition for the glasma fields in 
the initial stages of a heavy ion collision. A correlator of 
Wilson lines from the same classical fields, known as the 
dipole cross section, can be used to compute many inclusive and
exclusive observables in DIS.  
\end{abstract}

\section{Connection between small $x$ DIS and HIC: Wilson line}
The initial condition in a heavy ion collision (HIC) is determined by the
wavefunctions of the two colliding nuclei, parametrized by $Q^2$ and
$x$. As in any hadronic collision, the typical magnitudes of these
parameters can be estimated as $Q^2\sim \langle p_\perp \rangle^2$
and $x\sim  \langle p_\perp \rangle/\sqrt{s}$, where 
$\langle p_\perp \rangle$ is the typical transverse momentum of
the particles being produced, and $\sqrt{s}$ is the collision energy.
At relativistic energies, such as at RHIC and LHC, this means that 
the relevant domain for bulk particle production is at very
small $x$. Gluon brehmsstrahlung processes lead to an exponentially 
(in rapidity $y = \ln 1/x$) growing cascade of gluons in the wavefunction.
The number of gluons in the wavefunction grows as
$\ud N/\ud y \sim x^{-\lambda}$, where 
the phenomenologically observed value is $\lambda \sim 0.2 \dots 0.3$.
When the number of gluons grows large enough, eventually their 
phase space density becomes large, with occupation numbers
$\sim 1/\as$; in terms of the field strength this meas $A_\mu \sim 1/g$.
At this point the nonlinear terms in the QCD Lagrangian (think
of the two terms in the covariant derivative $\partial_\mu + i g A_\mu$)
become of the same order as the linear ones, and the dynamics becomes 
nonperturbative. Due to the nonlinear interactions the gluon number cannot
grow indefinitely, but it must \emph{saturate} at some $1/\as$ for gluons
with $p_\perp \lesssim \qs$, where  $\qs$ is the \emph{saturation scale}.
When $\qs^2 \sim x^{-\lambda}$ becomes large enough, $\as(\qs) \ll 1$
and the dynamics of these fields is \emph{classical}. 
This situation is most conveniently described using the effective theory
known as the Color Glass Condensate (CGC, \cite{Iancu:2003xm,*Weigert:2005us}),
where the large $x$ degrees of freedom are described as a classical
color current $J^\mu$ and the small $x$ gluons as classical fields that 
this current radiates: $[D_\mu,F^{\mu\nu}]= J^{\nu}$.

\begin{figure}
\centerline{
\resizebox{\textwidth}{!}{
\includegraphics[height=5cm,clip=true]{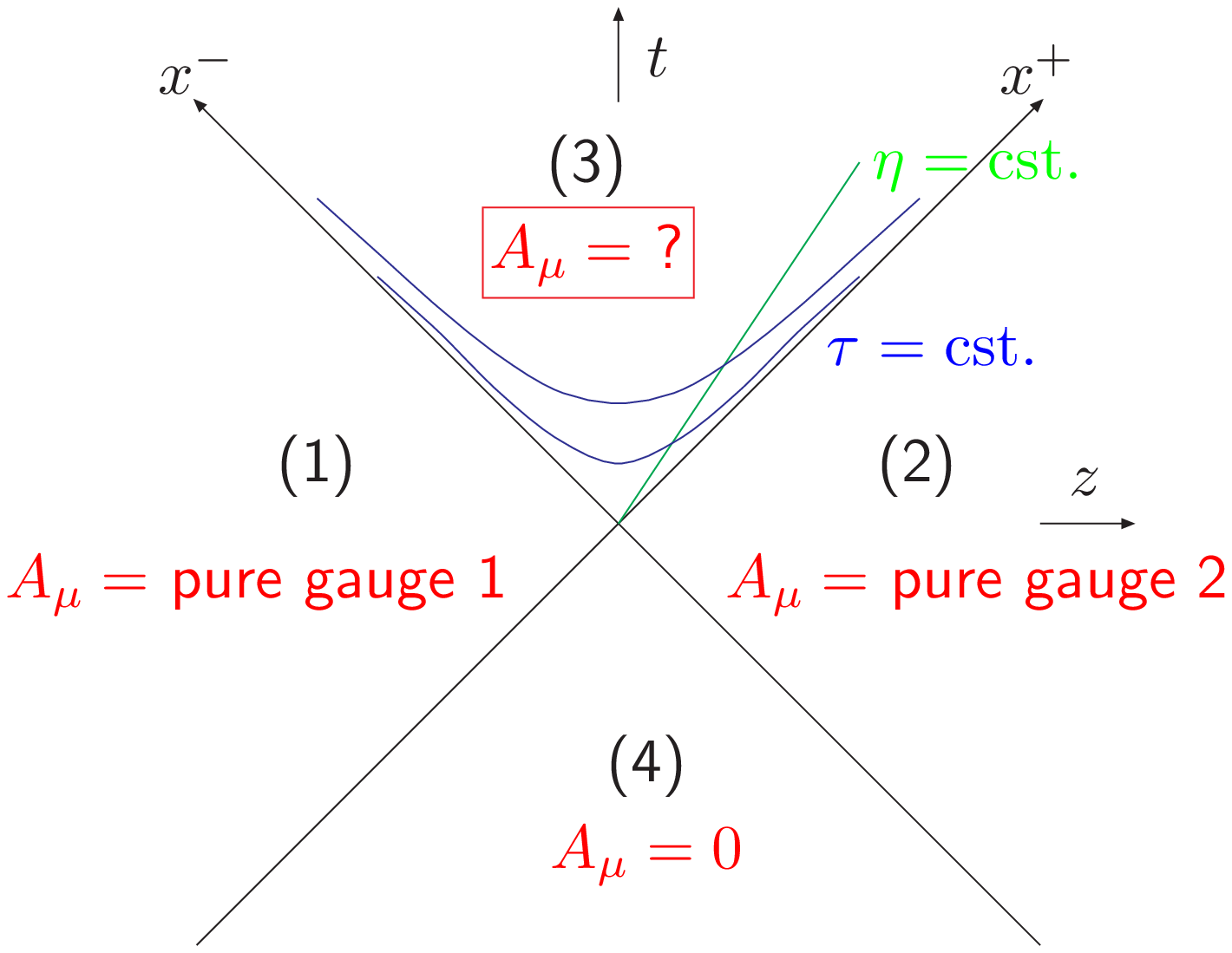}
\rule{1cm}{0pt}
\includegraphics[height=5cm,clip=true]{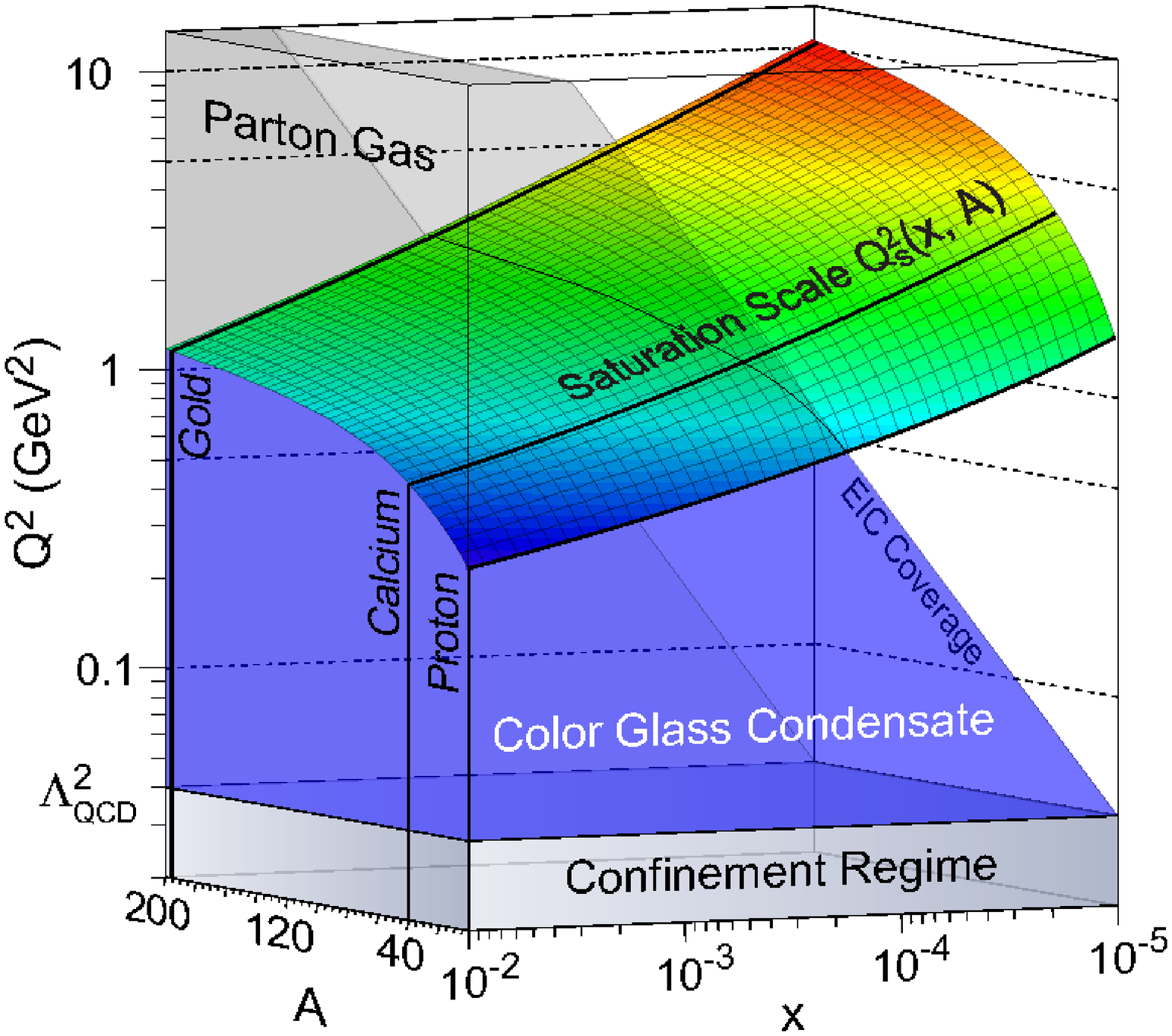}
}
}
\caption{Left: Spacetime structure of the field. In regions 
(1) and (2) there is a transverse pure gauge field~\nr{eq:pureg} with no energy 
density. In region (3) after the collision there is the glasma field.
Right: 
} \label{fig:spacet} \label{fig:qsa}
\end{figure}

Let us consider a hadron or a nucleus moving in the $+z$-direction. 
Its color current in the CGC formalism has only one large component, the one 
in the $+$-direction (recall that $x^\pm = (t \pm z)/\sqrt{2}$).
For a nucleus moving at high energy we can take the current to be independent
of the light cone time $x^-$ as $J^+ = \rho(\xt,x^-)$ with a very narrow,
$\delta$-functionlike support in $x^-$: 
$\rho(\xt,x^-) \sim \delta(x^-) \rho(\xt)$.
A simple solution for the equations of motion can be found as
$A^+ = \rho(\xt,x^-)/\nabt^2$; this is known  
as the covariant gauge solution.
In order to have a physical partonic
picture of the gluonic degrees of freedom it is necessary to gauge transform
this solution to the light cone gauge $A^+=0$. The gauge transformation
that achieves this is done with the \emph{Wilson line} constructed from
this gauge field: 
\begin{equation}\label{eq:defwline}
U(\xt)= \mathbb{P}\exp\left[i \int \ud x^- A^+\right].
\end{equation}
This results in a field with only transverse components:
\begin{equation}\label{eq:pureg}
A^i_\textnormal{(1,2)} = \frac{i}{g}
U_\textnormal{(1,2)}(\xt) 
\partial_i U_\textnormal{(1,2)}^\dag(\xt)
\end{equation}
for both of the colliding nuclei $(1,2)$ separately.
The intial condition for the ``glasma''~\cite{Lappi:2006fp}
fields at $\tau>0$ is given in terms of these pure 
gauges~\cite{Kovner:1995ts,*Kovner:1995ja}.
\begin{eqnarray}
\left. A^i\right|_{\tau=0} &=& A^i_\textnormal{(1)} + A^i_\textnormal{(2)} 
\\ \nonumber
\left. A^\eta\right|_{\tau=0} &=& \frac{ig}{2}
 [A^i_\textnormal{(1)} , A^i_\textnormal{(2)} ]
\end{eqnarray}
Inside the future light cone $\tau>0$ the field equations must 
be solved either numerically or in some approximation scheme.
The spacetime structure described here is illustrated in 
\fig\ref{fig:spacet} (left).
In the rest of this talk we shall be referring to the numerical 
``CYM'' (Classical Yang-Mills) 
computations~\cite{Krasnitz:2001qu,*Lappi:2003bi,*Krasnitz:2003jw}.

To see the connection to DIS it is convenient to consider the process in 
a Lorentz frame where the virtual photon has a large longitudinal momentum.
In the target rest frame 
(or more properly the ``dipole frame''~\cite{Mueller:2001fv} that does not
leave all the high energy evolution in the probe) the timescales of the
quantum fluctuations of the virtual photon are extremely slow. In order to 
interact with a hadronic target it must therefore split into a 
quark-antiquark pair already long before the scattering. This 
$q\bar{q}$-dipole  then interacts with the hadronic target with a 
scattering amplitude whose imagimary part is known as the ``dipole
cross section''. As typical hadronic scattering amplitudes at high energy, 
that of the dipole is almost purely imaginary, and we shall here neglect the
real part. The dipole cross section can be obtained from the quark propagator
in the gluonic background field of the target, which is quite naturally given
by the same Wilson line \nr{eq:defwline}~\cite{Buchmuller:1996xw}.
The dipole cross section (which, in general, is a function of
the size of the dipole $\rt$, the impact parameter $\bt$
and $x$) is the correlator of two Wilson lines 
\begin{equation}
\hat{\sigma}(\rt) = \int \ud^2 \bt 
\frac{1}{\nc}\left\langle 1 -
U^\dag\left(\bt + \frac{\rt}{2}\right)U\left(\bt - \frac{\rt}{2}\right) \right\rangle.
\end{equation}
For example, the total virtual photon cross section can be obtained by convoluting 
the dipole cross section with the virtual photon wavefunction which relates
 the $Q^2$ of the photon to the size of the dipole $r \sim 1/Q$:
\begin{equation}
\sigma^{\gamma^*p}_{L,T} =
\int \ud^2 \bt \int \ud^2 \rt  \int \ud z 
\left| \Psi^\gamma_{L,T}(Q^2,\rt,z) \right|^2 
\sigma_{\mathrm{dip}}(x,\rt,\bt).
\end{equation}
Fourier-transforming instead of simply integrating over the impact parameter dependence
gives access to the momentum transfer to the target in diffractive scattering.
The inclusive diffractive virtual photon cross section (really the elastic dipole-photon
cross section) is proportional to the square of the dipole cross section 
\begin{equation}
\frac{\sigma^{D,tot}_{L,T}}{\ud t} =
\frac{1}{16\pi}
\int \ud^2 \rt \int \ud z 
\left| \Psi^\gamma_{L,T}(Q^2,\rt,z) \right|^2 
\sigma^2_{\mathrm{dip}}(x,\rt,\boldsymbol{\Delta})
\end{equation}
and diffractive vector meson production can be obtained by projecting on the
virtual photon wavefunction 
\begin{equation}
\frac{\sigma^{D,V}_{L,T}}{\ud t} = 
\frac{1}{16\pi}
\left| \int \ud^2 \rt \int \ud z 
\left(\Psi^\gamma {\Psi^*}^V\right)_{L,T}  (Q^2,\rt,z) 
\sigma_{\mathrm{dip}}(x,\rt,\boldsymbol{\Delta}) \right|^2.
\end{equation}
The exclusive cross sections are proportional to the dipole cross section 
(the scattering
amplitude) squared, whereas the inclusive one depends on it linearly; this is due to 
the optical theorem and our approximation that the scattering amplitude is purely 
imaginary.
We shall now go on to discuss some recent applications of saturation
ideas to heavy ion collisions and DIS phenomenology, trying to stress
the unity of the approach between the two.

\begin{figure}
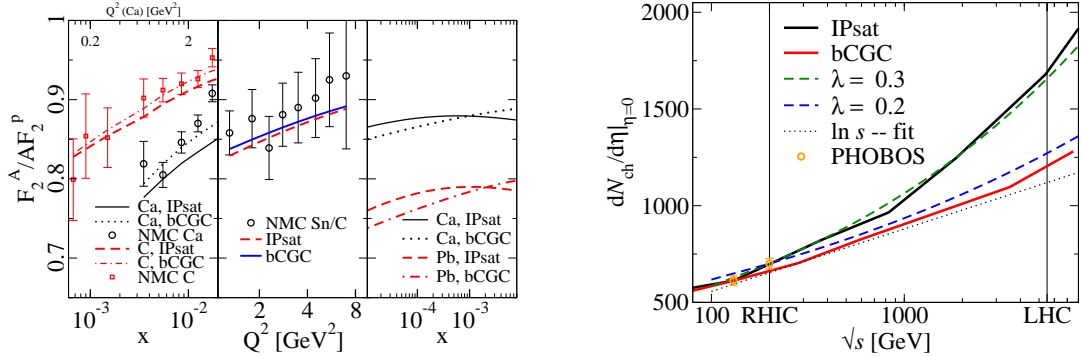

\centerline{
\resizebox{\textwidth}{!}{
\includegraphics[height=5cm,clip=true]{lappi_tuomas_eds09_fig2a.eps}
\rule{1cm}{0pt}
\includegraphics[height=5cm,clip=true]{lappi_tuomas_eds09_fig2b.eps}
}
}
\caption{Left: Comparison of the fit~\cite{Kowalski:2007rw} to
existing nuclear DIS data from NMC.
Right: Extrapolation of the gluon multiplicity to LHC energies,
from \cite{Lappi:2008eq}} \label{fig:multi}
\end{figure}

\section{Gluon multiplicity at RHIC and LHC AA collisions}

Ideally one would like to measure the value of $\qs$ in DIS 
experiments and use the resulting value as an independent 
input in calculations of the initial state of heavy ion 
collisions. In practice most of the exiting CYM computations
of the glasma fields have been performed in the MV
model~\cite{McLerran:1994ni,*McLerran:1994ka,*McLerran:1994vd}
in terms of the color charge density parameter 
$g^2\mu$ that parametrizes the fluctuations of the classical color 
currents $J^\pm$. One must therefore relate the values of $g^2\mu$
and $\qs$ in a consistent way. In practive this can be done
by computing the Wilson line correlator in the MV model, using exactly
the same numerical implementation of the model as in the CYM 
calculations, and extracting the correlation length 
$1/\qs$~\cite{Lappi:2007ku}. 

The other ingredient necessary in using the existing DIS data 
to calculate initial conditions for heavy ion collisions is 
the correct implementation of the nuclear geometry in 
extending the parametrization from protons to nuclei. In a ``Glauber''-like
formulation of essentially independent scatterings of the dipole on 
each of the nucleons this is a straightforward estimate, 
see e.g. Refs.~\cite{Kowalski:2007rw,Kowalski:2003hm}. 
A simple geometrical argument would give  the
estimate $\qsa^2 \approx 0.5 \qsp^2 A^{1/3}$, where the coefficient 
in front follows from the internucleon distance in a nucleus 
being smaller than the nucleon radius. The actual values
in the estimate of Ref.~\cite{Kowalski:2007rw} are shown in 
\fig\ref{fig:qsa} (right).
For other estimates of $\qs$ based on DIS data see
Ref.~\cite{Freund:2002ux,*Armesto:2004ud}.
Being really consistent with high energy evolution would require some 
further theoretical advances, since the approximation of independent
dipole-nucleon scatterings will break down during the evolution. 
In the infinite momentum frame this can be thought of as gluons
from different nucleons starting to interact with each other.

Combining these ingredients the CYM
calculations~\cite{Krasnitz:2001qu,*Lappi:2003bi,*Krasnitz:2003jw} 
of gluon production 
paint a fairly consistent picture of gluon production at RHIC energies.
 The estimated value $\qs \approx 1.2 \gev$ from HERA 
data~\cite{Kowalski:2007rw,Kowalski:2003hm} (corresponding to the MV
 model parameter $g^2 \mu \approx 2.1 \gev$~\cite{Lappi:2007ku})
gives a good description of existing nuclear DIS data from 
the NMC collaboration, see \fig\ref{fig:multi}  (left).
The same value
leads to $\frac{\ud N}{\ud y} \approx 1100$ gluons in the
 initial stage. Assuming a rapid thermalization and nearly 
ideal hydrodynamical evolution this is consistent with the
 observed $\sim 700$ charged ($\sim 1100$ total) particles
 produced in a unit of rapidity in central collisions. 

The gluon multiplicity is, across different 
parametrizations, to a very good 
approximation proportional to $\pi \ra^2 \qs^2/\as$. Thus 
the predictions for LHC collisions depend mostly
on the energy dependence of $\qs$. On this front there is 
perhaps more uncertainty than is generally 
acknowledged, the estimates for  
$\lambda= \ud \ln \qs^2 / \ud \ln 1/x$ varying between 
$\lambda= 0.29$~\cite{Golec-Biernat:1998js}
and $\lambda= 0.18$~\cite{Kowalski:2006hc} in fixed
 coupling fits to HERA data, with 
a running coupling solution of the BK equation 
giving something in between these 
values~\cite{Albacete:2004gw}. This dominates the 
uncertainty in predictions for the LHC multiplicity,
see \fig\ref{fig:multi} (right).

\section{Multiplicity distributions}

One very recent application of the CGC framework has been computing
the probability distribution of the number of gluons in the glasma~\cite{Gelis:2009wh}.
The dominant contributions to multiparticle correlations come from diagrams that
are disconnected for fixed sources and become connected only
after averaging over the color charge configurations. In other words,
the dominant correlations are those arising from resummed large logarithms
of the collision energy and are present already in the initial wavefunctions
of the colliding nuclei.

Working with the MV model Gaussian probability distribution
\begin{equation}
W[\rho] = \exp \left[ -\int \ud^2\xt \frac{\rho^a(\xt)\rho^a(\xt)}{g^4\mu^2} \right]
\end{equation}
computing the correlations in the linearized approximation
 is a simple combinatorial problem. 
The result can be expressed in 
terms of two parameters, the mean multiplicity $\bar{n}$, and a
parameter $k$ describing the width of the distribution.
The $q$'th factorial moment of the multiplicity ditribution is,
to leading order in $\as$, proportional to  $2^{q}(q-1)!$.
Explicitly, the connected parts of the moments
$m_q \equiv \langle N^q \rangle$
are
\begin{eqnarray}
 m_q &=& (q-1)! \,  k \left(\frac{\bar{n}}{k} \right)^q \textrm{ with }
\\
k &\approx&  \frac{  (\nc^2-1)   \qs^2 S_\perp  }{2\pi}
\\
\bar{n} &=&  f_N \frac{1}{\as} \qs^2 S_\perp.
\end{eqnarray}
These moments define a \emph{negative binomial} distribution with parameters  
$k$ and $\bar{n}$, which has been used 
as a phenomenological observation in high energy hadron and nuclear collisions
already for a long 
time~\cite{Arnison:1982rm,*Alner:1985zc,*Alner:1985rj,*Ansorge:1988fg,*Adler:2007fj,*Adare:2008ns}.
In terms of the glasma flux tube picture this result has a natural  interpretation.
The transverse area of a typical flux tube is $1/\qs^2$, and thus there are 
$\qs^2 S_\perp= N_{\textnormal{FT}}$  independent ones. Each of these radiates
particles independently into $\nc^2-1$ color states in a Bose-Einstein distribution
(see e.g.~\cite{Fukushima:2009er}). A sum of $k \approx N_{\textnormal{FT}} (\nc^2-1)$ 
independent Bose-Einstein-distributions is precisely 
equivalent to  a negative binomial 
distribution with parameter $k$.

\section{Inclusive nuclear diffraction at eRHIC and LHeC}

\begin{figure}
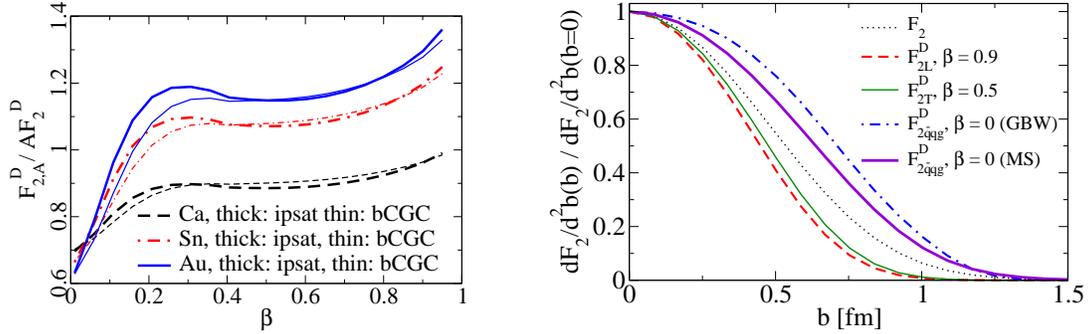

\centerline{
\resizebox{\textwidth}{!}{
\includegraphics[height=5cm,clip=true]{lappi_tuomas_eds09_fig3a.eps}
\rule{1cm}{0pt}
\includegraphics[height=5cm,clip=true]{lappi_tuomas_eds09_fig3b.eps}
}}
\caption{
Left: nuclear modification of the diffractive structure function, 
showing characteristic suppression at small $\beta$ (large mass of the
diffractive system) and enhancement at large $\beta$. 
From Ref.~\cite{Kowalski:2008sa}.
Right: dominant impact parameters for the different contributions
to the proton diffractive structure function.
} \label{fig:diff}
\end{figure}

The large fraction of diffractive events observed at HERA shows that
modern colliders are approaching the nonlinear regime of QCD, where gluon
saturation and unitarization effects become important. It should 
be possible to perform the same measurements in DIS off nuclei. 
There are plans for several
facilities capable of high energy nuclear DIS experiments, as the
EIC~\cite{Deshpande:2005wd} and LHeC~\cite{Dainton:2006wd} colliders.
Due to the difficulty in measuring an intact recoil nucleus deflected
by a small angle, diffractive eA collisions present an experimental
challenge. But if they are successful,
nuclear diffractive DIS (DDIS) would provide
a good test of our understanding of high energy QCD.
Measuring the momentum transfer $t$ in both coherent (nucleus stays intact)
and incoherent (nucleus breaks up into nucleons) would enable one to
go measure directly the transverse structure of the
gluonic degrees of freedom~\cite{Caldwell:2009ke} instead of the 
electric charge distribution that is measured in low energy experiments.
Figure~\ref{fig:diff} (left) demonstrates some expected results from such a
measurement. The diffractive structure function 
can be divided into different components according to the polarization
state of the virtual photon and the inclusion of higner Fock states
(e.g. $q\bar{q}g$ in addition to $q\bar{q}$) in the dipole wavefunction.
All of these have different dependences on the impact parameter
of the dipole-target collision (see \fig\ref{fig:diff} right), which stresses
the importance of having a detailed picture of the transverse geometry
of both the proton and the nucleus.

\section*{Acknowledgements}  
Numerous conversations with F.~Gelis, L.~McLerran and R.~Venugopalan are 
gratefully acknowledgements.
The author is supported by the Academy of Finland, contract 126604.

\bibliographystyle{h-physrev4mod2M}
\bibliography{spires}

\end{document}